\documentclass[onecolumn,10pt,a4paper,twoside]{IEEEtran}

\usepackage{amsmath, amssymb,graphicx,cite}
\usepackage{epsfig,amsfonts,subfigure}
\usepackage{graphicx,cite,amssymb,amsmath}
\usepackage{color}
\usepackage{grffile}

\usepackage[caption=false,font=footnotesize]{subfig}
\begin{document}
\graphicspath{{./Figures/}}
\title{On the Stability of Contention Resolution Diversity Slotted
ALOHA}
%
%
%

\author{Christian~Kissling
\thanks{C. Kissling is with the Institute of Communications and Navigation, German Aerospace Centre (DLR),
Oberpfaffenhofen, 82234 Germany e-mail: christian.kissling@dlr.de.
\newline This work was presented in part at the IEEE GLOBECOM 2011.
\newline This work is submitted for possible publication to the IEEE
Transactions on Communications. The IEEE copyright notices apply.
}
}

\maketitle



%
\IEEEpeerreviewmaketitle

\begin{abstract}
In this paper a Time Division Multiple Access (TDMA) based Random Access (RA) channel with Successive Interference Cancellation (SIC) is
considered for a finite user population and reliable retransmission
mechanism on the basis of Contention Resolution Diversity Slotted ALOHA (CRDSA). A general mathematical model
based on Markov Chains is derived which makes it possible to predict
the stability regions of SIC-RA channels, the expected
delays in equilibrium and the selection of parameters for a stable
channel configuration. Furthermore the model enables the estimation
of the average time before reaching instability. The presented model
is verified against simulations and numerical results are provided
for comparison of the stability of CRDSA versus the stability
of traditional Slotted ALOHA (SA). The presented results show that CRDSA
has not only a high gain over SA in terms of throughput but
also in its stability.
\end{abstract}

\section{Introduction}
While the application of RA techniques for data transmissions
is appealing in many application scenarios such as sensor networks,
signalling or unpredictable and bursty low duty cycle user traffic,
often concerns are expressed about the limitations in terms of
spectral efficiency and the risk of RA channel instability,
leading to a zero throughput and a correspondingly infinite
transmission delay. While recently significant improvements of the
spectral efficiency have been achieved by introducing SIC and
coding techniques (e.g. \cite{Casini2007}, \cite{Liva2010} and
\cite{Paolini2011}), the stability behaviour of these new schemes
has not been fully analyzed yet. In \cite{Kissling2011} a first
analysis of the stability of CRDSA as SIC representative
was done and a mathematical model for the prediction of the channel
stability was derived, which is used as baseline for the work
presented in this paper and therefore recalled in the following.

\begin{figure}[!ht]
    \centering
       \includegraphics[width=0.4\textwidth]{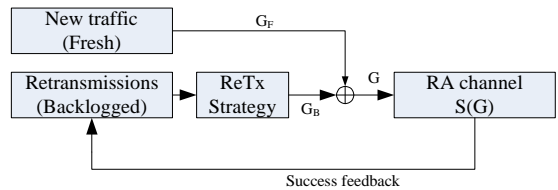}
    \caption{Feedback loop of a RA channel.}
    \label{FIG:FeedbackLoopRa}
\end{figure}

The source of channel instability in a RA channel is the
natural occurrence of collisions among the packet transmission and
the presence of mechanisms which attempt the retransmission of lost
packets. In principle collided packets could be simply discarded,
but doing so would adversely affect the Quality of Service (QoS) experienced by the
user or may be entirely unacceptable for critical signalling
information, such as log-on messages. RA schemes thus usually
attempt to retransmit the lost packets, either until they are
successfully received or until a maximum number of retransmissions
has been reached. In order to make retransmissions possible, the
users need to receive feedback whether their transmission attempt
was successful, e.g., by means of acknowledgements. The
instantaneous throughput of the RA channel $S(G)$ is then
dependent on the total load $G$, being the sum of the load due to
new transmissions $G_F$ and the load due to retransmissions $G_B$.
In this sense the RA channel forms a feedback loop as is
illustrated in Fig. \ref{FIG:FeedbackLoopRa}. It is an inherent
property of closed-loop feedback systems, that the feedback can lead
to amplifying self-excitation. Here this results in an increase of
the overall load due to the additional retransmissions.

The throughput curves of ALOHA, SA \cite{D.Bertsekas1992}, and
CRDSA \cite{Casini2007} all have in common that for increasing
load $G$ the throughput $S(G)$ first increases until reaching a
maximum throughput $S_{max}$. For further increasing load, $S(G)$
decreases again and asymptotically approaches zero.

If due to retransmission attempts of lost packets the total load
exceeds a critical threshold, then even more packets experience a
collision and get lost, resulting in an even higher retransmission
load. In the end the channel is driven into total saturation in the
area of having very high load and very low throughput. To reduce
this amplification effect a retransmission strategy is used, which
shall limit the load due to retransmissions and reduce the risk of
getting more collisions (see Fig. \ref{FIG:FeedbackLoopRa}). Many
different retransmission strategies that try to achieve this goal
are known from literature. In \cite{Kleinrock1973} the selection of
the time of retransmission with uniform probability within a
parameterizable interval $t \in [0,\, ... ,\,K]$ is proposed. In
\cite{D.Bertsekas1992} a strategy is described where the decision
for a retransmission attempt is taken with a probability $p_r$ in
every slot (for SA) resulting in a geometric distribution. In
\cite{Metcalfe1976} the selection of the retransmission time from an
interval, which grows exponentially with every collision (Binary
Exponential Backoff), is proposed. Finally the so called
\emph{splitting algorithms} (see e.g., \cite{D.Bertsekas1992},
\cite{Capetanakis1979}) iteratively split the set of collided users
into two sets and stabilize the system this way. Furthermore two
different types of user population are distinguished, \emph{finite}
and \emph{infinite} user populations. For a finite user population,
every user that experienced a collision is backlogged, which means
that he is not generating any new traffic until the collided packet
has been successfully transmitted. The infinite user population on
the other hand refers to either an infinite number of users or a
finite number of users that generate new traffic independently of
whether another retransmission is still pending or
not.\footnote{Generating a new transmission in addition to a
retransmission can be also seen as two users, one retransmitting,
one transmitting new data. Since the generation of new transmissions
is not bounded, the user population can also grow to infinity.}
While some retransmission strategies assume a visibility of the
channel activities by all users, here we assume that every user has
no instant visibility of other users activity (as is the case in
satellite systems with directive links and long propagation delays)
and only receives feedback about the success of his own transmission
attempt from the receiving end system. The retransmission mechanisms
using a uniform and geometric retransmission probability have in
common that the probability of retransmission is a fixed parameter
and does not change dynamically. For the binary exponential backoff
and tree splitting algorithm, the actual retransmission probability
may change over time dependent on the situation. In the remainder of
this paper the focus is on a geometrically distributed
retransmission mechanism, since it was shown in \cite{Lam1974} that
the channel performance of SA is mainly dependent on the
average retransmission delay and largely independent of the
retransmission probability distribution.

\section{Review of SIC-based Random Access Techniques}

\begin{figure}[t]
    \centering
        \includegraphics[width=0.4\textwidth]{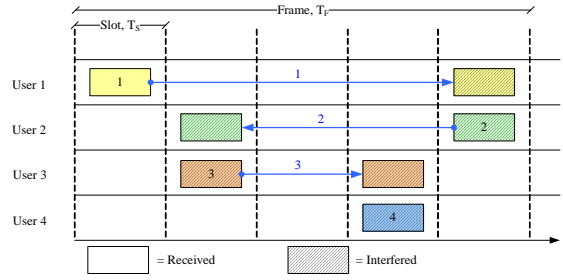}
    \caption{SIC principle of CRDSA. Numbers at arrows indicate the step in the iteration process.}
    \label{FIG:CrdsaPrinciple}
\end{figure}

Over the last years the recently regained popularity of RA
schemes resulted in the definition of new RA protocols. In
particular a recent enhancement of the SA protocol, named
CRDSA \cite{Casini2007}, \cite{RioHerrero2009}, using SIC
techniques over a set of slots (denoted frame) to improve the
throughput and Packet Loss Rate (PLR) behaviour of SA, has been studied
showing an impressive gain over SA increasing the maximum
throughput from $S_{max,SA}=0.36 \frac{pkt}{slot}$ to
$S_{max,CRDSA}=0.55 \frac{pkt}{slot}$. Up to now however the
consequences for the system stability of this new access scheme have
not been analyzed yet. The fundamental concept of CRDSA is to
generate a replica burst for every transmission burst within a set
of $N_S$ slots, called frame, see also Fig.
\ref{FIG:CrdsaPrinciple}. While the generation of a redundant copy
of a burst is similar to previous proposals such as Diversity Slotted ALOHA (DSA)
\cite{Choudhury1983}, the fundamental difference here is that every
burst contains a pointer to the location of its replica. In case a
clean replica arrives, meaning that the burst could be decoded and
received successfully, the channel is estimated from it and the
interference that this burst introduces to other users is removed
for all replica-burst locations.

In the example in Fig. \ref{FIG:CrdsaPrinciple} the first burst of
user 1 is received successfully since not interfered. As consequence
of the SIC process, the interference that the replica of user 1
introduces to the second burst of user 2 is removed so that this
burst of user 2 can be decoded in the next round. This process is
then iteratively repeated. In the example in Fig.
\ref{FIG:CrdsaPrinciple} all replicas can be recovered this way.

\subsection{Characterization of the Packet Loss Rate in
CRDSA}\label{SEC:CharacterizationPLR}

For classical SA, the necessary condition to have a successful
reception is that only a single transmission must occur in a
timeslot, otherwise the burst is lost. Let us denote by $M$ the
total user population of the system and $p_0$ the probability that a
user attempts a transmission in a time slot, then the probability
that a user successfully receives a packet gets $p_{succ,SA} = p_0
\cdot (1-p_0)^{M-1}$. Increasing the overall number of users $M \to
\infty$, the totally transmitted packets can be modeled as Poisson
process with arrival rate $\lambda$ \cite{Abramson1977}. The
probability for a successful transmission then results in the well
known equation $p_{succ,SA} = \lambda \cdot T_p \cdot e^{-\lambda
\cdot T_p}$, whereas $T_p$ denotes the slot duration. This simple
closed form expression is conveniently suited to describe the
throughput surfaces, which are used for the stability investigation
done e.g., by Kleinrock \cite{Kleinrock1975}. The preconditions in
CRDSA are however different due to the iterative SIC
process. As was shown by Liva in \cite{Liva2010} and
\cite{Liva2010a}, the SIC process can be interpreted as an
erasure decoding process in a bipartite graph, such as for Low Density Parity Check Codes (LDPC)
codes \cite{Gallager1963}. For this purpose, every slot in a frame
is represented as a \emph{sum node} and every transmitted burst by a
\emph{burst node}. The edges in the graph then connect the burst
nodes to the sum nodes. In \cite{Liva2010} an expression for the
average erasure probabilities for every iteration are derived for
the asymptotic case of infinitely long frames, resulting in an upper
bound of the achievable throughput. An expression for neither the
exact nor the average erasure probabilities in a non-asymptotic case
with finite frame lengths however can be expressed accurately by
these bounds or another closed form expression.
For this reason the stability analysis in this work relies on
simulated CRDSA packet success probabilities and throughput for
the case of having one additional replica (degree $d=2$), a frame
consisting of $N_S=100$ slots and a limitation of the number of
SIC iterations to $I_{max}=10$. The presented framework is
however flexible to be used as well for other configurations of
CRDSA, always requiring only that the average throughput curve
is known.

\subsection{Stability Definition}
The issue of stability in RA systems was already identified in
the very early days of the ALOHA proposal. Abramson
\cite{Abramson1970} and Roberts \cite{Roberts1975} both addressed
this issue for plain ALOHA. After the evolution of ALOHA towards
SA, many publications have dealt with the investigation of the
stability behaviour of SA, for instance \cite{D.Bertsekas1992},
\cite{Gallager1985}, \cite{Abramson1977} and \cite{Kleinrock1975}.
Stability is commonly defined as the ability of a system to maintain
equilibrium or return to the initial state after experiencing a
distortion. In the context of RA, the term \emph{stability} is
used in different ways in literature. In the definition given by
Abramson in \cite{Abramson1970}, the ALOHA channel was defined
instable if the average number of retransmissions becomes unbounded.
Within \cite{D.Bertsekas1992} a channel was defined stable if the
expected delay per packet is finite. Kleinrock defined in
\cite{Kleinrock1975} a channel as stable if the SA equilibrium
contour (i.e., throughput is equal to the channel input rate) is
nontangentially intersected by the load line in exactly one place.
In the strict mathematical definition of stability of autonomous
systems, this corresponds to a sufficient condition for a global
equilibrium point. In the terminology used by Kleinrock, a SA
channel is instable if the load line intersects the equilibrium
contour in more than one point. In the mathematical sense also then
the system can have a locally stable equilibrium point, so the
definition of stability by Kleinrock refers to the criterion of
having a single globally stable equilibrium point. In the remainder
of this work, the definitions given in \cite{Kleinrock1975} are
followed also here, meaning that a channel is denoted as stable if
it has a single globally stable equilibrium point and instable
otherwise.

\section{Stability in CRDSA}\label{SEC:StabilityInCrdsa}
Within this section, the derivation of a Markov model for a finite
user population is described and the mathematical formulations for
throughput and drift are derived, which form the core of the
stability framework presented afterwards. This section concludes
with a stability analysis for a representative CRDSA
configuration.

Let the RA channel under consideration be populated by a total
of $M$ users (finite user population). Every user  resides either in
a so called \emph{fresh (F)} state or \emph{backlogged (B)} state.
In the beginning all $M$ users are in state \emph{F}. Every user in
state \emph{F} attempts a new transmission in the current frame with
probability $p_0$. It is further assumed that all users receive
feedback about the success of their transmission at the end of a
frame. In case the transmission attempt was successful, the user
remains in state \emph{F}. In case a packet is lost, the user enters
state \emph{B}. A user in state \emph{B} attempts a retransmission
of the lost packet with probability $p_r$ in the current frame. In
case the retransmission is successful the user then returns to state
\emph{F}, otherwise the user remains in state \emph{B}. Let
$X_\alpha^l$ denote the number of users in state $\alpha \in \left
\{ F,B \right \}$ in frame $l$, then the discrete-time Markov chain
can be fully described by either $X_B^l$ or $X_F^l$, since both are
connected by $X_F^l = M - X_B^l$. In the following $X_B^l$ is chosen
as the Markov state variable. Given the initial state $X_B^0=0$ and
the state transition probability $P(x'| x)$, which is the
probability to move within one frame from backlog state $x$ to state
$x'$, the Markov chain is then fully described. One major difference
to the SA analysis done by Kleinrock is that the backlog state
for SA can at maximum decrease by 1 user per slot (otherwise
there would be a collision), while the backlog state $X_B$ for
CRDSA can decrease by $[1,\,\ldots,\, X_B]$ in a frame. Since
no closed form expression for the success probability of a user in
CRDSA is known in literature, the probability
$q_{d,N_s,I_{\text{max}}}(\tau, \upsilon)$ is introduced, which is
the probability that out of $\tau$ users who attempt a transmission
in the frame exactly $\upsilon$ users are successful. The success
probability $q$ is dependent on the CRDSA configuration,
consisting of the repetition degree $d$, the number of slots in the
frame $N_s$ and the maximum number of iterations $I_{max}$. Here,
this probability was derived numerically by simulations and
averaging over the results for every offered load $G$. For sake of
simplicity, the subscripts will be omitted in the following, using
$q(\tau,\upsilon)$. When changing state, let $\upsilon_l$ be a
random variable denoting the number of successful transmissions in
frame $l$, $\varphi_l$ a random variable denoting the number of
fresh transmission attempts in the frame and $\rho_l$ the number of
retransmission attempts in a frame.

Let $FS_l$ denote the number of fresh users, which transmit
successfully in frame $l$. Let $FU_l$ be the number of fresh users
who attempted a transmission but were unsuccessful. In the same way,
$BS_l$ denominates all backlogged users who attempt a retransmission
and were successful and $BU_l$ those backlogged users, whose
retransmission attempt was unsuccessful. The following equations
\eqref{eqn1}-\eqref{eqn3} can be derived: \footnote{It should be
noted that idle users have no relevance here since they neither
change the size of the sets $X_F$ and $X_B$ nor do they generate
load which impacts the transmission performance.}
\begin{eqnarray}
    \varphi_l = FS_l + FU_l, \label{eqn1}\\
    \rho_l = BS_l + BU_l, \label{eqn2}\\
    \upsilon_l = FS_l + BS_l. \label{eqn3}
\end{eqnarray}
The joint probability mass function, conditioned on state $X^l=x_B$
is then given by Eq. \eqref{eqn4}.
\begin{eqnarray}
    \lefteqn{ P(\varphi,\rho,\upsilon | x_B) = {} } \nonumber \\
    & & {} = \binom{x_F}{\varphi} p_0^\varphi (1-p_0)^{x_F-\varphi} \cdot
    \binom{x_B}{\rho} p_r^{\,\rho} (1-p_r)^{x_B-\rho} \nonumber \\
    & & {} \qquad \cdot q(\varphi+\rho,\upsilon) = \nonumber \\
    & & {}  = \binom{M-x_B}{\varphi} p_0^\varphi (1-p_0)^{M-x_B-\varphi} \nonumber \\
    & & {} \qquad \cdot \binom{x_B}{\rho} p_r^{\,\rho} (1-p_r)^{x_B-\rho} \nonumber \\
    & & {} \qquad \cdot q(\varphi+\rho,\upsilon).   \label{eqn4}
\end{eqnarray}
With Eqs. \eqref{eqn1}-\eqref{eqn3} the change in number of
backlogged users $\Delta x_B = x_{B,l+1} - x_{B,l} = x_B' - x_B = FU
- BS$ can be easily reformulated into:
\begin{equation}
    \upsilon = \varphi + x_B - x_B' \label{eqn7}
\end{equation}
The state transition probability $P(x'_B | x_B)$ can then be
formulated by combining \eqref{eqn4} and \eqref{eqn7} into
\eqref{eqn5}:
\begin{eqnarray}
    \lefteqn{ P(x'_B | x_B) = {} } \nonumber \\
    & & {} = \sum_{\varphi,\rho} P(\varphi,\rho, \varphi + x_B - x'_B | x_B) = \nonumber \\
    & & {} = \sum_{\varphi,\rho} \binom{M-x_B}{\varphi} p_0^\varphi (1-p_0)^{M-x_B-\varphi}  \nonumber \\
    & & {} \qquad \cdot \, \binom{x_B}{\rho} p_r^{\,\rho}  (1-p_r)^{x_B-\rho}  \nonumber \\
    & & {} \qquad \cdot q(\varphi+\rho,\varphi+x_B-x'_B). \label{eqn5}
\end{eqnarray}
With \eqref{eqn5} in principle the entire Markov chain can be
described with all its transition probabilities. In practice the
computational cost of computing all transition probabilities is
however enormous, mainly due to the nested summations over a large
range of possible values for $\varphi$ and $\rho$. To avoid this
computational complexity, the stability analysis in the following
makes use of a drift analysis, in reminiscence of
\cite{Kleinrock1975} and \cite{Murali1997}. The change of the
backlog state forms a differential equation, whereas the drift
corresponds to the change of the state variable $d_B=\frac{d
x_B}{dt}$. For the drift analysis the change in backlog $x_B$ over
time is analyzed in the following and the stability of the
equilibrium points is computed by using the tools known from
differential calculus. In the style of \cite{Murali1997} and
\cite{Carleial1975}, the drift is here defined as the expectation of
the change of the backlog state $X_B^l$ frame by frame as given by:
\begin{eqnarray}
\lefteqn{ d(x_B) = d_B = E \left \{ X_B^{l+1} - X_B^{l} \, | \, X_B^l \right \} = {} } \nonumber \\
    & & {} = \sum_{x'_B} (x'_B-x_B) \cdot P(x'_B \, | \, x_B).
\end{eqnarray}
whereas $E\{.\}$ denotes the expectation value $E_{x|y} \{ f(x) \} =
\sum f(x) \cdot p(x|y)$.

With \eqref{eqn7} and \eqref{eqn5} this can be
reformulated into:
\begin{eqnarray}
\lefteqn{ d(x_B)=\sum_{\varphi,\rho,\upsilon} (\varphi-\upsilon) \cdot P(x'_B \, | \, x_B) = {} } \nonumber \\
    & & {} = \sum_{\varphi,\rho,\upsilon} (\varphi-\upsilon) \cdot \sum_{\varphi,\rho} P(\varphi,\rho,\upsilon \, | \, x_B) = \nonumber \\
    & & {} = \sum_{\varphi,\rho,\upsilon} (\varphi-\upsilon) \cdot P(\varphi,\rho,\upsilon \, | \, x_B) =  \nonumber \\
    & & {} = E \left \{ \Phi \right \} - E \left \{ \Upsilon \right \}.
\end{eqnarray}
whereas $\Phi$ denotes the random variable taking the values
$\varphi$ and $\Upsilon$ the random variable taking the values
$\upsilon$.

From \eqref{eqn4} it is clear that $\varphi$ is binomial distributed
so:
\begin{equation}
    E \{ \Phi \} = (M-x_B) \cdot p_0.
    \label{eqn11}
\end{equation}

\begin{figure}[!t]
    \centering
        \includegraphics[width=0.5\textwidth]{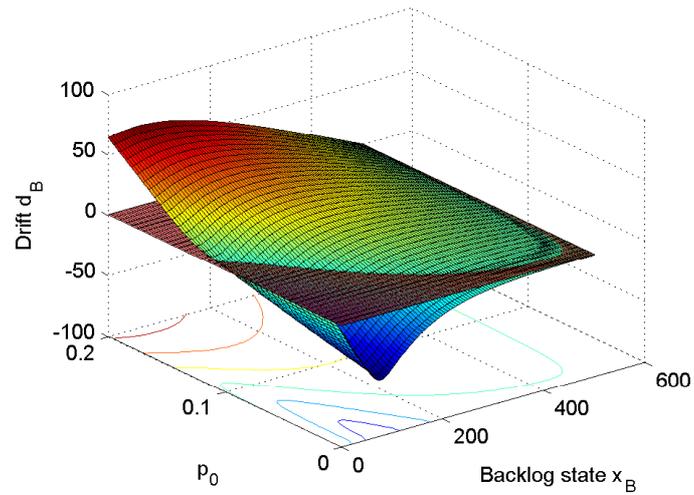}
    \caption{Drift backlog surface for $M=500$, $p_r=0.78$.}
    \label{FIG:DriftBacklSurf_1}
\end{figure}

The second expectation value $E\{ \Upsilon \}$ is related to the
throughput $S(x_B)$ of the system by \eqref{eqn6}, i.e., the
expected number of successful packets per slot in frame $l$:
\begin{eqnarray}
\lefteqn{ S(x_B) = \frac{1}{N_S} E \{ \Upsilon \} = {} } \nonumber \\
    & & {} = \frac{1}{N_S} \cdot \sum_{\varphi,\rho,\upsilon} \upsilon \cdot P(\varphi,\rho,\upsilon \, | \, x_B) = \nonumber \\
    & & {} = \frac{1}{N_S} \cdot \sum_{\varphi,\rho,\upsilon} \upsilon \binom{M-x_B}{\varphi} p_0^\varphi (1-p_0)^{M-x_B-\varphi} \nonumber \\
    & & {} \cdot \binom{x_B}{\rho}p_r^{\,\rho} (1-p_r)^{x_B-\rho} \cdot q(\varphi+\rho,\upsilon). \label{eqn6}
\end{eqnarray}
The expected throughput $S(x_B)$ can also be expressed via the
average success probability $\overline{P}_s(x)$, i.e. the
probability of a successful transmission in a frame when attempting
$x$ transmissions:
\begin{equation}
    S(x_B) = \left [ (M-x_B) p_0 + x_B p_r \right ] \cdot \overline{P}_s ((M-x_B) p_0 + x_B p_r). \label{eqn10}
\end{equation}
With \eqref{eqn11}, \eqref{eqn6} and \eqref{eqn10} the drift $d_B$
becomes:
\begin{equation}
    d_B = (M-x_B) \cdot p_0 - N_S \cdot S(x_B). \label{eqn9}
\end{equation}
With \eqref{eqn9} it is now possible to fully describe the stability
of the CRDSA system for the case of having a user population
$M$, a probability $p_0$ of fresh users generating new packets and a
retransmission trial probability of $p_r$. Intuitively, the drift
represents the tendency of the system to change over time and gives
the direction of change of the backlog size. This means that for
positive drifts the size of the backlog tends to increase by $d_B$
(i.e., more users experience lost packets and get backlogged). For
negative drifts, the length of backlog decreases, which means that
backlogged users successfully retransmit and get fresh again. A
drift of 0 corresponds to an equilibrium point, which may be locally
stable or instable. Fig. \ref{FIG:DriftBacklSurf_1} shows the
dynamics of the channel with the drift-backlog surface for the
scenario $M=500$, $p_r=0.78$ and for varying $p_0$. The surface can
be classified into three different areas: In the first area for $0
\leq p_0 \leq 0.01$ the drift-backlog-surface does not intersect the
zero-drift plane and gets the tangent plane for $p_0=0.01$. In the
second area for $ 0.01 < p_0 \leq 0.11$ the drift-backlog-surface
intersects the zero-drift plane in three equilibrium points. In the
third area for $0.11 < p_0 \leq 1$ the drift-backlog-surface
intersects the zero-drift plane in a single equilibrium point which
is located at the saturation point where all or almost all $M$ users
are backlogged.

\begin{figure}[!t]
    \centering
        \includegraphics[width=0.5\textwidth]{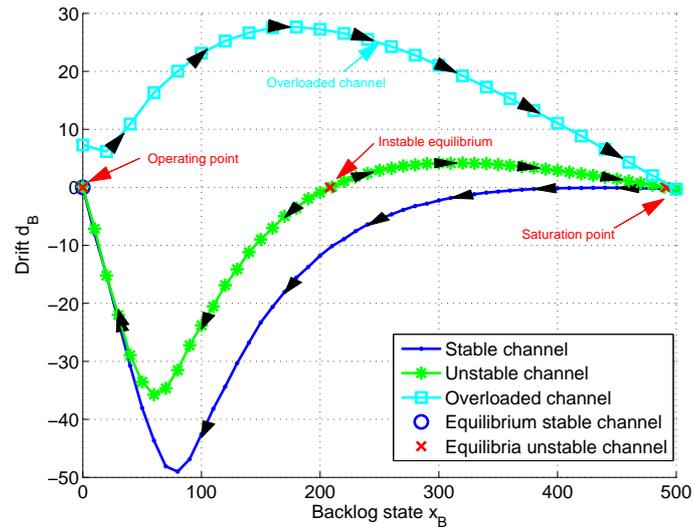}
    \caption{Backlog drift for a stable channel with $M=500$, $p_r=0.78$ and $p_0=0.01$, an instable channel with $M=500$, $p_r=0.78$ and $p_0=0.04$ and
    an overloaded channel with $M=500$, $p_r=0.78$ and $p_0=0.12$.}
    \label{FIG:DriftCut_1_combined}
\end{figure}

Fig. \ref{FIG:DriftCut_1_combined} shows the backlog drift of the
three areas for three representative values of $p_0$, i.e. the
intersection of the drift-backlog surface from Fig.
\ref{FIG:DriftBacklSurf_1} with the planes $p_0=\left\{0.01, 0.04,
0.12 \right\}$.

As can be seen here, the drift for the stable configuration ($p_0
\leq 0.01$) is always negative independent of the backlog state
$x_B$ and approaches asymptotically a drift $d_B=0$, which means
that the system always shows the tendency to lower the current
backlog state until reaching the initial state. There is thus only
one equilibrium point (globally stable) close to the initial state.

For the instable configuration ($0.01 < p_0=0.04 \leq 0.11$) it can
be seen that after the initial equilibrium point (locally stable)
and the following area of negative drift (up to $x_B \approx 209$) a
second, locally instable equilibrium point is reached at $x_B=209$.
When reaching this point the system can either fall back into the
negative drift region for $x_B < 209$ or enter the region of
positive drift $x_B> 209$. In the latter case the positive drift
means that any movement to a higher backlog state (which is a
consequence of the positive drift) results in an accelerated
increase in number of backlogged users. This behavior then persists
until reaching the third and final equilibrium point (locally
stable) at $x_B \approx 500=M$. In this third equilibrium point now
all or almost all $M$ users are backlogged and the system has
reached the point of maximum load and minimum throughput. In case
some of the users get unbacklogged, the drift is anyway positive and
drives the channel back into the saturation. While there is a low
probability that the channel returns in the high throughput region,
the probability is fairly small and it can be expected that a very
long time passes before this happens. For the overloaded
configuration $0.11 \leq p_0=0.12 \leq 1.0$ the drift-backlog
surface intersects the zero drift plane only once at the saturation
point where all or almost all $M$ users are backlogged. Since there
are no equilibria before and the drift is always positive, it can be
expected that the system moves straight towards the saturation point
after being started. An instable system may remain for some time in
the desirable high throughput region of the operating point before
getting instable and entering the low throughput region around the
saturation point. For an overloaded configuration the channel moves
directly to the saturation point.

From this observation the conclusion can be drawn that for a given
system configuration $\Omega=\{ d, N_s, I_{max} \}$ the maximum
traffic generation probability $p_0$ for which the system is still
always stable is the one resulting in a drift contour which
intersects the straight line $d_B=0$ at most once (Fig.
\ref{FIG:DriftCut_1_combined}). The resulting single equilibrium
point is then locally and globally stable. For all other cases the
channel is instable (e.g., instable configuration with $p_0=0.04$ in
Fig. \ref{FIG:DriftCut_1_combined}), meaning that earlier or later
the backlog will increase into the total saturation point. If the
single equilibrium point coincides with the saturation point, the
system is overloaded. While it is - mathematically speaking - also
stable in this scenario (locally and globally stable equilibrium) it
is in total saturation with very low throughput and high delay.
Since an operating point in this region is not viable for a
communication system, in the remainder of this paper stable refers
only to having one equilibrium point in the high throughput region.

With this framework, it is now possible to predict the stability of
a channel with a certain set of parameters or to derive a set of
parameters for which the channel is guaranteed to be stable. In
\cite{Kissling2011} the validity of this model was verified against
simulations in different scenarios.

\section{Average Delay}
From the stability model defined in the previous section it can be
observed that the stability of a system with fixed $p_0$ benefits
from a reduction of the retransmission probability $p_r$. Or in
other words a configuration which is instable can always be
stabilized by decreasing $p_r$. This comes however at the cost of a
higher delay since reducing $p_r$ means increasing the average time
before attempting a retransmission. On the one hand a low average
delay (i.e. requiring $p_r$ to be as high as possible) is important
for achieving a good QoS perception for the user. On the other
hand remaining stable is important for user satisfaction as well,
since an instable system will be driven into total saturation with
asymptotically zero throughput and infinite delay. For a stable
configuration however it is beneficial to have a $p_r$ as small as
possible. The retransmission probability $p_r$ is thus a design
parameter which can be optimized to achieve a delay as low as
possible while being selected high enough to ensure a stable system
operation. For this reason it is important to derive an analytical
framework that makes it possible to compute the expected delay for a
given system configuration in order to find the optimum choice for
the design parameter $p_r$, e.g., to minimize the delay while
remaining stable, but also for optimizing the maximum allowable
packet generation probability $p_0$ or the maximum allowable user
population $M$ which is treated in section
\ref{SEC:STABILITY_COMPARISON_FOR_STABLE_CHANNELS}. While the
stability model derived in the previous sections provides the
mathematical framework to derive the overall set of parameters for
which the RA is stable, this section deals with the computation
of the expected delay for any set of parameters. \newline

 The analysis of SA in \cite{Kleinrock1975} followed the fundamental
principles of Markov theory and derives the expected delay $D_b$ via
Little's theorem. According to this well known theorem, the average
number of packets in a queuing system in stable conditions is the
product of the packet arrival rate and the average dwell time in the
queue. Applied to the stability analysis, the expected dwell time in
the queue corresponds to the transmission delay $D_b$ of every
packet (i.e. the time the packet remains in the channel until it is
successfully received). The average number of packets in the channel
is given by the expected backlog length $\overline N$ since every
backlogged user has one pending transmission. In equilibrium the
traffic arrival rate is equal to the serving rate, or in other words
the channel throughput $S_0$ is equal to the offered load $G_0$.
Following this analogy, the expected delay in a random access
channel computes with Little's theorem to:
\begin{equation}
D_b = \frac{\overline{N}}{S_{out}} \label{EQN:AvgDelayLittle_1}
\end{equation}
whereas
\begin{displaymath}
S_{out} = \sum_{n=0}^{M} S_{out}(n,p_0) \cdot P_n
\end{displaymath}
and $P_n$ is the probability of being in state $n$. Similarly the
expected backlog length $\overline{N}$ can be computed as:
\begin{displaymath}
\overline{N} = \sum_{n=1}^{N} n \cdot P_n.
\end{displaymath}

As it has been shown in \cite{Kleinrock1975} by numerical
simulations, the values for $S_{out}$ and $\overline{N}$ can be
closely approximated by the equilibrium point throughput $S_0$ and
backlog state $n_0$, i.e. $S_{out} \approx S_0$ and $\overline{N}
\approx n_0$.
With this and \eqref{EQN:AvgDelayLittle_1} the expected delay $D_b$
gets:
\begin{equation}
    D_b = \frac{n_0}{S_0} \label{EQN:AvgDelayLittle_2}.
\end{equation}

In order to show that the approximations of $S_{out}$ and
$\overline{N}$ claimed by Kleinrock for SA are also valid in
the case of CRDSA, the theoretical expected $D_b$ and the
measured delay $D_b^{sim}$ have been compared for a representative
CRDSA configuration $\Psi = \{M,p_0,p_r,d,N_s,I_{max} \} =
\{200, 0.9, \frac{1}{60}, 2, 100, 10 \}$.

The channel is stable in this configuration with an equilibrium
point at $n_0=149.06$ and an average throughput of $S_0=0.46$. With
\eqref{EQN:AvgDelayLittle_2} the expected delay gets
$D_b=324.04\,\text{slots}=3.2404\,\text{frames}$. The average delay
obtained by simulations is $D_b^{sim}= 328.64\,\text{slots} =
3.2864\,\text{frames}$, which is fairly close to $D_b$ and thus
confirms firstly that the approximation for $S_{out}=S_0$ and
$\overline{N}=n_0$ are also valid in the case of CRDSA, and
secondly that the presented framework is suitable to estimate the
average delay for a given channel configuration $\Psi$.

\section{Stability Comparison of SA and CRDSA for Stable
Channels}\label{SEC:STABILITY_COMPARISON_FOR_STABLE_CHANNELS} With
the ability to compute the expected delay for a given configuration
$\Psi$, the stability of CRDSA can now be compared to the
stability of SA. The stability of SA was deeply
investigated in \cite{Kleinrock1975}. The comparison of the two
stabilities is of particular interest since CRDSA offers much
higher throughput rates, also for higher offered traffic loads but
the question arises whether this gain comes at the cost of lower
stability, or not. For comparing the two RA schemes it needs to
be ensured that the conditions are comparable. For the stability and
performance of the RA schemes a tradeoff exists between the
total user population\footnote{Here only finite user population
scenarios are considered} $M$, probability of traffic generation
$p_0$ for unbacklogged users (user activity) and retransmission
probability $p_r$ for backlogged users. As it was explained earlier,
the RA channel for a finite user population can always be
stabilized by choosing a low enough retransmission probability
$p_r$. The selection of $p_r$ on the other hand impacts the delay,
as shown before, e.g., a lower value of $p_r$ will have a positive
impact on the stability of the system but results in longer delays.
For comparing the SA and CRDSA stability the following
optimization criteria can be chosen now:
\begin{enumerate}
\item Minimize the average delay $D_b$ for fixed user population $M$ and
fixed traffic generation probability $p_0$
\item Maximize the size of the user population $M$ for fixed $p_0$
and average delay $D_b$
\item Maximize the supported traffic generation probability $p_0$ for
fixed $M$ and $D_b$
\end{enumerate}
In the following analysis of these three criteria, CRDSA
configurations are specified by the set $\Psi$ whereas the SA
configurations are denoted by the set $\Xi=\{M,p_0,p_r\}$. It should
be noted that the traffic generation probability $p_0$ refers to the
probability of generating a packet in a transmission frame for
CRDSA whereas it refers to the probability of generating a
packet in a time slot for SA. To ensure a fair comparison among
the two, i.e., having the same overall traffic generation, $p_0$ for
SA is chosen to \eqref{EQN:p0_SA_CRDSA_RELATION} in the
following.
\begin{equation}
p_0^{SA} = \frac{p_0^{CRDSA}}{N_s} \label{EQN:p0_SA_CRDSA_RELATION}
\end{equation}

\subsection{Comparison of Achievable Delay
$D_b$}\label{SEC:STABLE_DELAY_COMPARISON} %
\begin{figure}[!ht]
    \centering
        \includegraphics[width=0.5\textwidth]{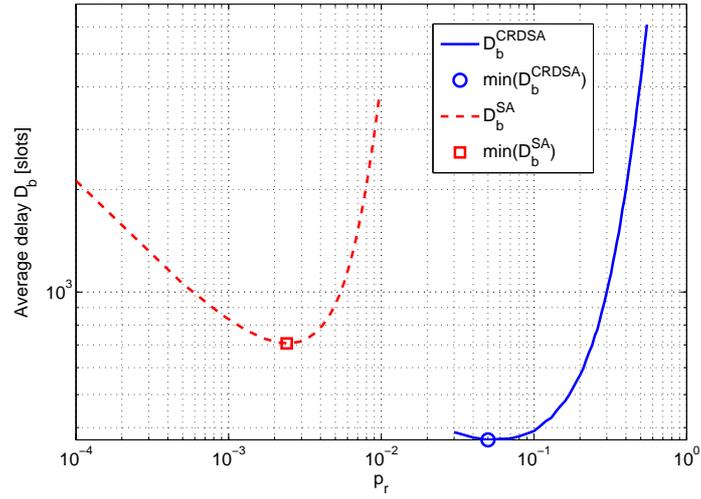}
    \caption{Comparison of minimum achievable delay for SA and CRDSA and $\Psi' = \{M=400,p_0=0.263,p_r,d=2,N_S=100,I_{max}=10\}$ and
    $\Xi'=\{M=400,p_0=2.63 \cdot 10^{-3},p_r\}$.}
    \label{FIG:Db_400_0263_005_fig1}
\end{figure}

The design parameter $p_r$ impacts the stability of the system as
well as the resulting average delay. For a given user population $M$
and traffic generation probability $p_0$ this forms an optimization
problem of selecting the optimum $p_r^*$ which is low enough to
guarantee a stable operation of the channel, while it should be as
high as possible at the same time to provide a low average delay.
This optimization problem can be formulated for CRDSA in the
following way:
\begin{equation}
p_r^* = \underset{p_r \in [0, \dots, 1]}{\operatorname{argmin}}\,D_b(\Psi,p_r).
\end{equation}
Fig. \ref{FIG:Db_400_0263_005_fig1} illustrates this optimization
problem. As it can be seen, the argument resulting in the minimum
achievable delay for CRDSA and configuration $\Psi'=
\{M=400,p_0=0.263,p_r,d=2,N_s=100,I_{max}=10\}$ gets:
\begin{displaymath}
    p_{r,CRDSA}^* = \underset{p_r \in [0 \dots
    1]}{\operatorname{argmin}}\,D_b(\Psi', p_r) = 5 \cdot 10^{-2}
\end{displaymath}
and the resulting minimum average delay computes to:
\begin{displaymath}
D_{b,\Psi'}^{min}(p_r=p_{r,CRDSA}^* = 5 \cdot 10^{-2})=
3.68\,\text{frames} \equiv 368\, \text{slots}.
\end{displaymath}

For SA and comparable configuration $\Xi'=\{M=400,p_0=2.63
\cdot 10^{-3},p_r\}$ the optimization for the minimum achievable
delay results in the optimum retransmission probability
$p_{r,SA}^*=2.5 \cdot 10^{-3}$ and an average delay of:
\begin{displaymath}
D_{b,\Xi}^{min}(p_r=p_{r,SA}^*=2.5 \cdot 10^{-3})= 707\,
\text{slots} \equiv 7.07\,\text{frames}
\end{displaymath}

While naturally the gain in terms of delay of the two different
schemes changes with the other configuration parameters in $\Psi$
and $\Xi$, the results above show that in the given configuration
CRDSA can save $48\%$ of the delay compared to SA in same
conditions and guaranteeing a stable channel.

\subsection{Comparison of Supported User
Population}\label{SEC:STABLE_POPULATION_COMPARISON} The second
optimization criterion is to determine the maximum user population
$M$ which can be supported with the same average transmission delay
$D_b$ while guaranteeing the stability of the channel. Finding the
maximum user population $M^*$ for achieving an average delay $D_b^0$
forms an implicit optimization problem of $D_b(p_r,M)$ with side
condition \eqref{EQN:Side_Condition_1}:
\begin{equation}
    g(p_r,M)=D_b(p_r,M)-D_b^0=0. \label{EQN:Side_Condition_1}
\end{equation}

The solution of this optimization problem can be easily found with a
Lagrange auxiliary function \eqref{EQN:Lagrange_M}:

\begin{equation}
    L(p_r,M,\lambda)=D_b(p_r,M)+\lambda [D_b(p_r,M)-D_b^0]. \label{EQN:Lagrange_M}
\end{equation}

\begin{figure}[!ht]
    \centering
        \includegraphics[width=0.5\textwidth]{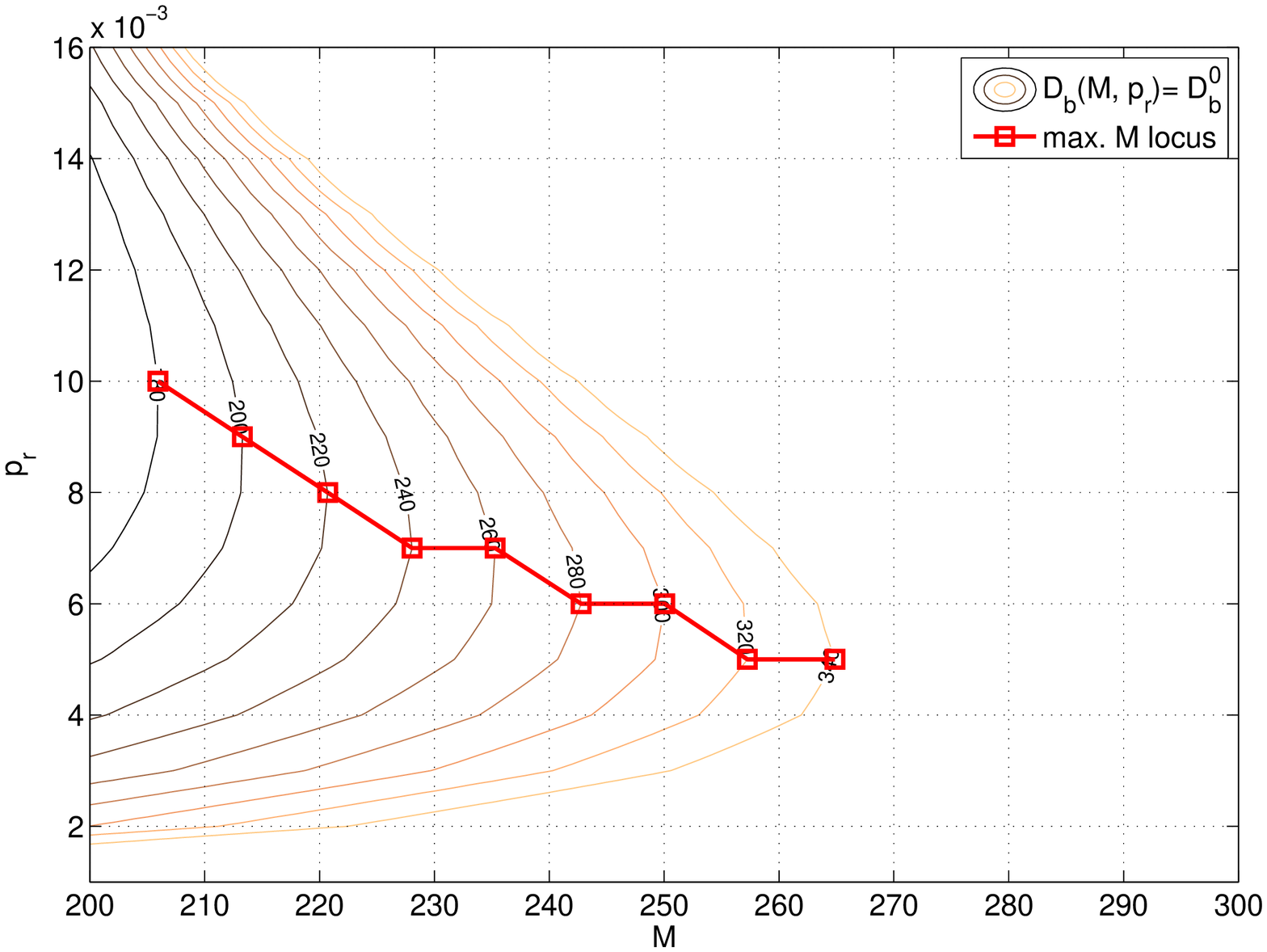}
    \caption{Locus for $D_b(M,p_r)=D_b^0$ with $D_b^0 \in [180 \dots 340]\,\text{slots}$ for SA and
    locus of max. supported user population $M(D_b^0)$.}
    \label{FIG:Contour_M_pr_p0_000263_SA}
\end{figure}

The maximum supported user population $M^*$ for retransmission
probability $p_r^*$ is then given simply by solving the set of
equations:
\begin{displaymath}
\nabla L(p_r, M,\lambda) = \left (
\begin{array}{l}
\frac{\partial L(p_r,M,\lambda)}{\partial p_r} \\
\frac{\partial L(p_r,M,\lambda)}{\partial M} \\
\frac{\partial L(p_r,M,\lambda)}{\partial \lambda}
\end{array}
\right ) = \underline{0}.
\end{displaymath}

resulting in the locus of tuples $(M^*(D_b^0);p_r^*(D_b^0))$ shown
in Fig. \ref{FIG:Contour_M_pr_p0_000263_SA} for SA with
$\Xi=\{M,2.63 \cdot 10^{-3},p_r\}$ and for different values of
$D_b^0$.

 As it can be seen, the maximum user population, which can be
 supported at a maximum delay of $D_b^0=300\,\text{slots}$ gets $M^*_{SA}=250$ for an optimum $p_{r,SA}^*=6\cdot
10^{-3}$.

\begin{figure}[!ht]
    \centering
        \includegraphics[width=0.5\textwidth]{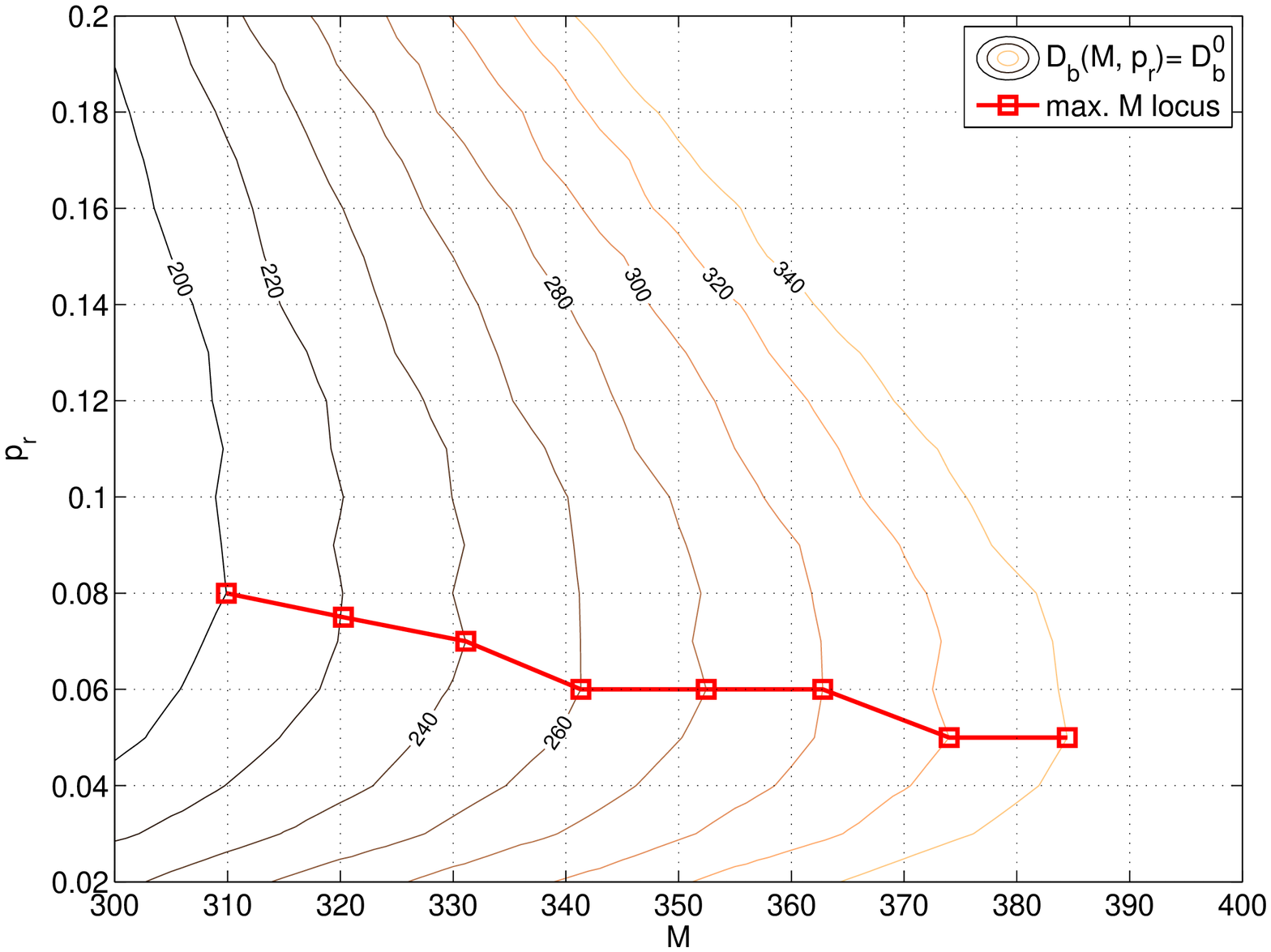}
    \caption{Locus for $D_b(M,p_r)=D_b^0$ with $D_b^0 \in [ 200 \dots 340]\,\text{slots}$ for CRDSA
    and locus of max. supported user population $M(D_b^0)$.}

    \label{FIG:Contour_M_pr_p0_0263_CRDSA}
\end{figure}

Fig. \ref{FIG:Contour_M_pr_p0_0263_CRDSA} shows the solution of the
same optimization problem for CRDSA. The traffic generation
probability was set to the equivalent value $p_0^{CRDSA}=2.63\cdot
10^{-1}$ in order to get the comparable traffic generation
probability as in SA, resulting in the configuration
$\Psi=\{M,2.63 \cdot 10^{-1},p_r,2,100,10\}$. As it can be seen
here, the joint optimization results in $M^*_{CRDSA}=363$ for a
retransmission probability $p_{r,CRDSA}^*=0.06$.

The comparison for this configuration shows that CRDSA can
support 45\% more users than SA while achieving the same
average delay and being also guaranteed stable.

\subsection{Comparison of Supported Traffic Generation
Probability}\label{SEC:STABLE_TRAFFIC_GENERATION_COMPARISON} 
\begin{figure}[!ht]
    \centering
        \includegraphics[width=0.5\textwidth]{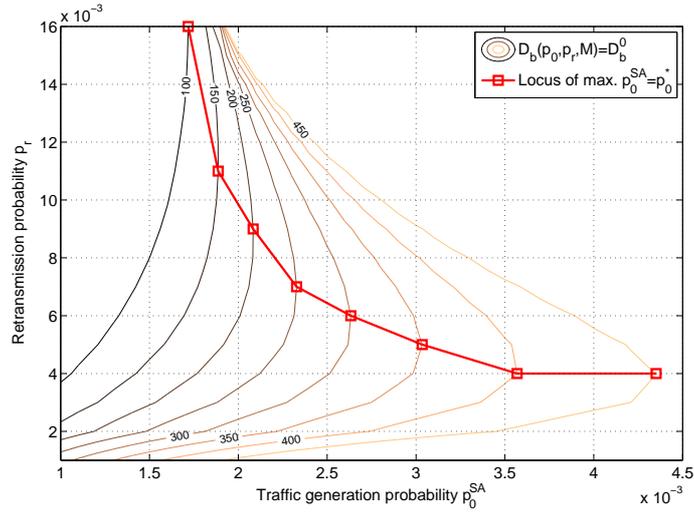}
    \caption{Joint optimization of $p_0$ and $p_r$ to achieve $D_b^0\,\text{slots}$ for SA with
    $\Xi=\{250, p_0, p_r\}$.}
    \label{FIG:Delay_Contour_p0_pr_Db_300_M_250_SA}
\end{figure}

\begin{figure}[!ht]
    \centering
        \includegraphics[width=0.5\textwidth]{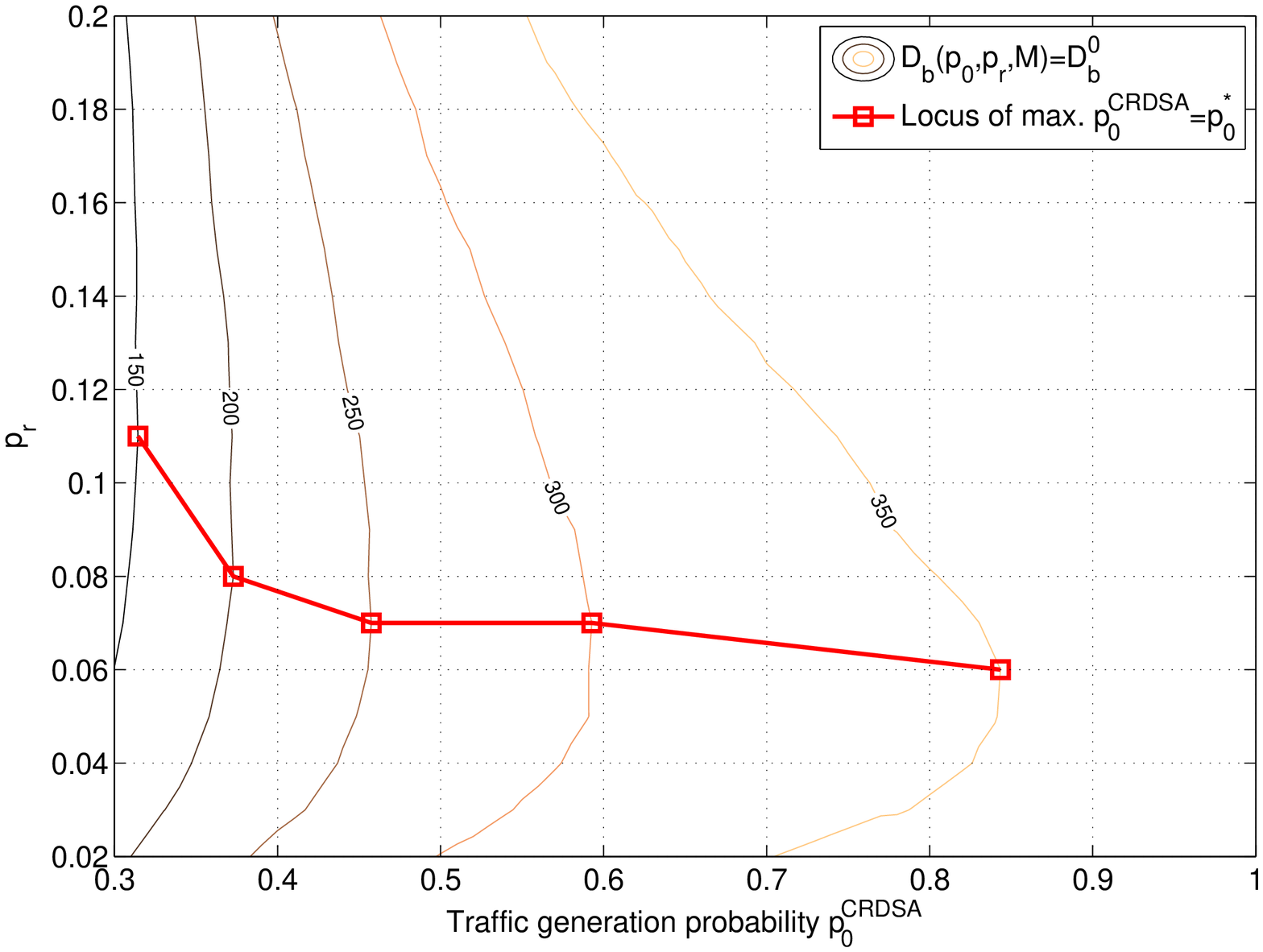}
    \caption{Locus for $D_b(p_0,p_r)=D_b^0$ with $D_b^0 \in [150 \dots 350]\,\text{slots}$ for CRDSA and locus of max. supported
    user population $M(D_b^0)$.}
    \label{FIG:Contour_p0_pr_M_250_CRDSA}
\end{figure}

For the
third optimization criterion, the user population is fixed together
with the average delay to be achieved while guaranteeing at the same
time that the channel remains stable. The optimization problem here
is very similar to the previous one in section
\ref{SEC:STABLE_POPULATION_COMPARISON} and consists in finding the
retransmission probability $p_r^*$ for which the the traffic
generation probability $p_0^*$ is maximized for given user
population $M$. Defining the Lagrange auxiliary function:
\eqref{EQN:Lagrange_p0}
\begin{equation}
    L(p_r,p_0,\lambda)=D_b(p_r,p_0)+\lambda [D_b(p_r,p_0)-D_b^0] \label{EQN:Lagrange_p0}
\end{equation}
and solving the set of equations given by:
\begin{displaymath}
    \nabla \,L(p_r,p_0,\lambda) = \underline{0}
\end{displaymath}
provides the locus of optimum tuples $(p_0^*,p_r^*)$ shown in Fig.
\ref{FIG:Delay_Contour_p0_pr_Db_300_M_250_SA} for SA and in
Fig. \ref{FIG:Contour_p0_pr_M_250_CRDSA} for CRDSA and
different $D_b^0$.

As it can be seen by comparing SA and CRDSA for e.g.
$D_b^0=350\,\text{slots}$, the traffic generation probability
supported by CRDSA is with $p_{0,CRDSA}^*=0.84$ a factor 2.8
higher than the one for SA with $p_{0,SA}^*=3 \cdot 10^{-3}$.
CRDSA thus allows users to generate traffic with a 2.8 times
higher traffic generation probability than SA.

\section{Average time before failure for
CRDSA}\label{SEC:INSTABLE_TIME_BEFORE_FAILURE} The previous
sections were focused on the investigation of stable channels. In
many application scenarios instability may be acceptable if the time
before getting instable is only sufficiently high. In stable channel
conditions, the performance comparison of RA schemes could be
done by comparing the minimum achievable delay $D_b$, the maximum
number of supported users $M$ or the maximum traffic generation
probability $p_0$. But for an instable channel configuration, these
criteria do not apply anymore. In an instable channel the operating
point will sooner or later reach the locally stable but undesired
equilibrium point $n_s$ in the low throughput region, which can also
coincide with the total saturation point, where all users are
backlogged and the delay $D_b$ grows to infinity. Also the maximum
user population $M$ or maximum traffic generation probability $p_0$
are no suitable measures. In an instable channel, $M$ and/or $p_0$
can grow arbitrarily while the channel will always remain instable.
What changes is the time to reach $n_s$ which will be shorter with
growing $M$ and $p_0$. With this in mind, the average time before
the channel enters the instable region for the first time can be
used as a suitable measure for comparing the behaviour of different
RA schemes in instability.\\
Once the undesired operating point $n_s$ in the high load/low
throughput area is entered, the channel may remain there potentially
for a very long time (unless it is being reset). There is only a
very small, but non-zero, probability to get out of this undesired
operating point which depends on the configuration $\Psi$. As
explained and shown in section \ref{SEC:StabilityInCrdsa} an
instable system has three operating points, two of them locally
stable and one locally instable. Among the two locally stable ones,
one resides in the low load area (desired operating point) whereas
the other one resides in the high-load/low throughput region (i.e.
high number or all users backlogged). When in the locally instable
operating point, the system has a chance to fall back into the
desired region but the same chance to enter the undesired region,
ending up in the low throughput operating point. In the SA
analysis done by Kleinrock \cite{Kleinrock1975}, this instable
operating point is also denoted as the critical system state $n_c$.
A measure for comparing the stability of different RA channels
is then the average time before the critical state $n_c$ is reached
for the first time, assuming further that the system will fall into
the low throughput region, once $n_c$ is reached. In the Markov
chain representation, the state $n_c+1=n_u$ is modeled as an
absorbing state in order to simplify the analytical analysis. It
should be noted that this is clearly only a model since in a real
system the probability of leaving the high backlog state is
non-zero, while it gets zero when using an absorbing state. In this
work the focus is only to derive the time until the system is
entering the instable state for the first time without looking at
the time until it would leave the instability region again. The
average first entry time $\overline{T}_i$ into state $i$ can be
expressed recursively by
\begin{equation}
    \overline{T}_i = 1 + \sum_{j=0}^{n_c} p_{ij} \cdot \overline{T}_j \label{EQN:FET_1}
\end{equation}
whereas $p_{ij}$ denotes the state transition probability from state
$i$ to state $j$
\begin{eqnarray}
    p_{ij} &=& Prob \left [ N^{t+1}=j | N^t = i \right ] = p_{x_B,x_B'} = \nonumber\\
    &=& \sum_{s,t} \binom{M-x_B}{\varphi} \cdot p_0^\varphi \cdot (1-p_0)^{M-x_B-\varphi} \cdot \nonumber\\
    &&  \binom{x_B}{\rho} \cdot p_r^{\,\rho} \cdot (1-p_r)^{x_B-\rho} \cdot \nonumber\\
    && q(\varphi+\rho, \varphi+x_B-x'_B) \label{EQN:p_ij_Markov_Model}
\end{eqnarray}

For the computation of the First Entry Times (FET), it is now of interest to know
the average time until reaching the critical state $n_c$ for the
first time when starting from the initial state $x_b^0=0$, i.e.,
$\overline{T}_0^{n_c}$.
 The recursive formulation in \eqref{EQN:FET_1} yields a set of linear
 equations:
\begin{displaymath}
\left [
\begin{array}{c}
    \overline{T}_0 \\
    \overline{T}_1 \\
    \vdots \\
    \overline{T}_{n_c}\\
\end{array}
\right ] = 
\left [
\begin{array}{c}
        1\\
        1\\
        \vdots\\
        1\\
\end{array}
\right ] + \left [
\begin{array}{cccc}
    p_{00} & p_{01} & \ldots & p_{0{n_c}} \\
    p_{10} &        &        &             \\
    \vdots &        &        &              \\
    p_{{n_c}0}&     &        & p_{{n_c}{n_c}}\\
\end{array}
\right ] \cdot 
\left [
\begin{array}{c}
    \overline{T}_0 \\
    \overline{T}_1 \\
    \vdots \\
    \overline{T}_{n_c}\\
\end{array}
\right ] \nonumber \\
\end{displaymath}
which can be expressed in matrix vector notation by Eq.
\eqref{EQN:FET_MatrixVector_1}
\begin{equation}
\mathbf{t} = \mathbf{e} + \mathbf{P} \cdot \mathbf{t} \label{EQN:FET_MatrixVector_1}
\end{equation}
whereas $\mathbf{e}$ is the unity vector. The vector of interest
with all the FETs for every state $i$ gets then
\begin{equation}
\mathbf{t} = \left ( \mathbf{I}-\mathbf{P} \right ) ^{-1} \cdot \mathbf{e} \label{EQN:FET_MatrixVector_2}
\end{equation}
with $\mathbf{I}$ being the identity matrix and $\mathbf{e}$ the
unity vector. For the stability measure of the channel the entry of
interest is the first entry in $\mathbf{t}$ which represents the
time $\overline{T}_0^{n_c}$ to reach $n_c$ starting from
the initial state $x_B=0$.%
\subsection{Validation of the
Model}\label{SEC:INSTABLE_MODEL_VALIDATION} 
\begin{figure}[!ht]
    \centering
        \includegraphics[width=0.5\textwidth]{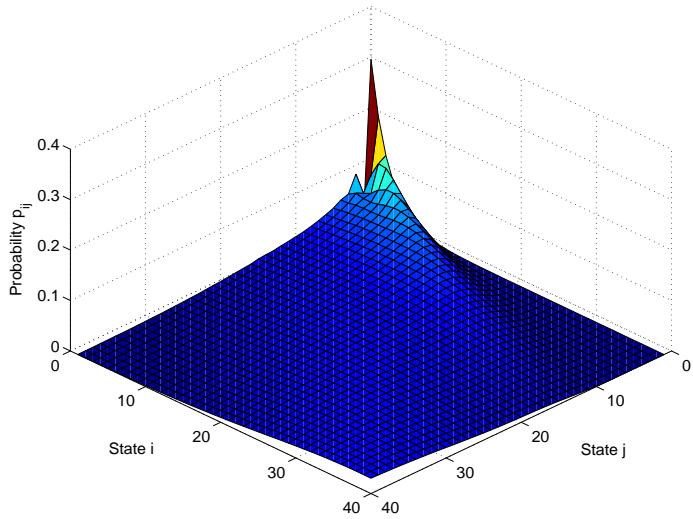}
    \caption{Markov state transition probabilities $p_{ij}$ for $\Psi_0$.}
    \label{FIG:MM_M300_p0_019_pr_07_3D_Computed}
\end{figure}

\begin{figure}[!ht]
    \centering
        \includegraphics[width=0.5\textwidth]{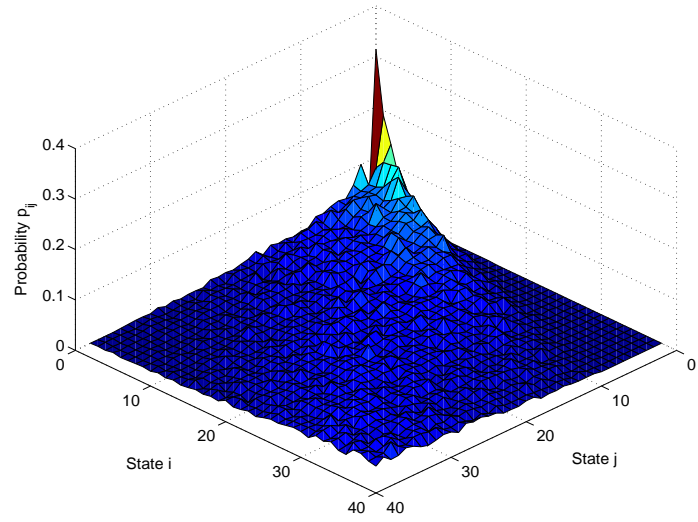}
    \caption{Simulated Markov state transition probabilities $p_{ij}$ for $\Psi_0$ (1000 runs).}
    \label{FIG:MM_M300_p0_019_pr_07_3D_Simulated}
\end{figure}

In order to illustrate
the validity of the derived model, the markov state transition
matrix and the FET is computed for a representative example
here. The considered scenario is $\Psi_0 = \{300 ,\, 0.19 ,\, 0.7
,\, 2 ,\, 100 ,\, 10\}$ which was chosen to result in an instable
CRDSA channel with equilibrium points at $n_0=20$ (locally
stable desired operating point), $n_c=40$ (locally instable
equilibrium) and $n_s=247$ (locally stable undesired operating
point). Since $n_u = n_c+1$ is an absorbing state it is sufficient
to compute the Markov state transition probabilities in the range of
states from $[0 \dots n_u]$. Fig.
\ref{FIG:MM_M300_p0_019_pr_07_3D_Computed} and
\ref{FIG:MM_M300_p0_019_pr_07_3D_Simulated} show the computed and
simulated Markov state transition matrix $\mathbf{P}$ for $\Psi_0$.

As it can be seen from the two graphs, the transition probabilities
resulting from the simulations match very well with the ones derived
by numerical computation.\\%
By solving the set of linear equations from
\eqref{EQN:FET_MatrixVector_2}, the computed FET time in this
example results in $FET_{comp}=14.57 \, \text{frames}$, where the
average simulative FET reaches $FET_{sim}=13.59 \,
\text{frames}$, which is very close to the expected FET derived
by computation.

\section{FET comparison between CRDSA and SA
    \label{SEC:INSTABLE_FET_COMPARISON}}
For a fair comparison between the FET of CRDSA and
SA, configurations need to be selected which have the same
initial conditions, i.e., the same user population $M$ and traffic
generation probability $p_0$. The average delay $D_b$ cannot be used
here since the average delay $D_b$ is infinite for an instable
channel by definition as the channel will enter the saturation point
with close to zero throughput.

A difficulty here consists in the fact that for an instable
CRDSA configuration, the SA channel is getting overloaded.

On the other hand an instable SA configuration for which three
equilibria exist results in a CRDSA configuration which is
stable so no FET can be computed.

\begin{figure}[!ht]
    \centering
        \includegraphics[width=0.5\textwidth]{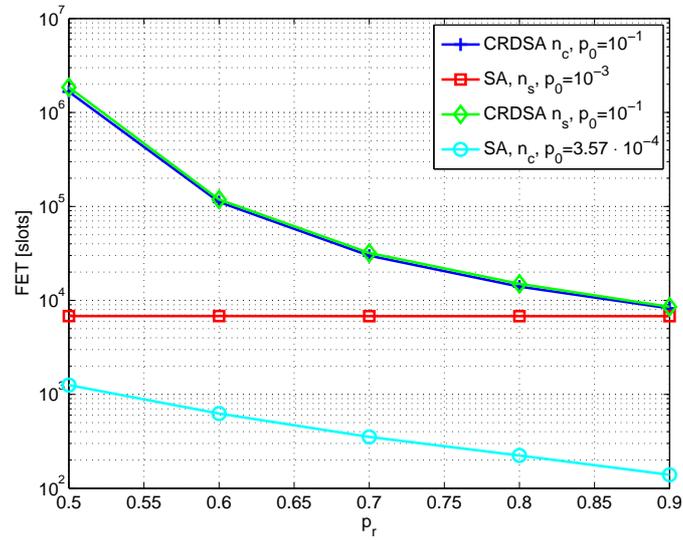}
    \caption{FET times for CRDSA and SA.}
    \label{FIG:FET_TIME_COMPARISON_1}
\end{figure}

Fig. \ref{FIG:FET_TIME_COMPARISON_1} shows the FET times for
different configurations of CRDSA and SA for a user
population of $M=500$.

As it can be seen, the FET for CRDSA and
$p_0^{CRDSA}=10^{-1}$ is up to a factor 20 higher than for the
equivalent SA configuration. It should be noted that SA is
already in overload for a $p_0^{SA}=10^{-3}$. For this reason the
SA FET curve does not show the time until reaching the
critical state $n_c$ but the time until reaching the saturation
point $n_s$ instead.

Fig. \ref{FIG:FET_TIME_COMPARISON_1} furthermore shows the FET
curve for CRDSA until reaching the saturation point $n_s$. The
FETs for this curve are slightly higher than for $n_c$ as could
be expected. This result also confirms the assumption to compute the
FET times by modeling $n_c$ as an absorbing state instead of
computing the full Markov chain up to $n_s$, since once $n_c$ is
reached also $n_s$ is reached very fast. The SA curve in Fig.
\ref{FIG:FET_TIME_COMPARISON_1} could arise the impression that the
FET curve is flat and has a qualitatively different shape than
the CRDSA curve. This is actually not the case and the
$\overline{T}_0^{n_c}$ for $p_r=0.5$ is indeed higher than for
$p_r=0.9$ with a value of
$\overline{T}_0^{n_c}(p_r=0.5)=6809.04\,\text{slots}$ and
$\overline{T}_0^{n_c}(p_r=0.9)=6798.45\,\text{slots}$. In this
configuration the SA channel is already so overloaded that also
a large decrease of the retransmission probability $p_r$ does not
have a significant impact anymore. Once the traffic generation
probability $p_0^{SA}$ is lowered, the impact of $p_r$ gets more
visible as it can be seen from the last curve in Fig.
\ref{FIG:FET_TIME_COMPARISON_1}, which was computed for
$p_0^{SA}=3.57\cdot 10^{-4}$. Also for this lower $p_0^{SA}$ the
$\overline{T}_0^{n_c}$ is much lower than for CRDSA with a
higher $p_0^{CRDSA}=10^{-1}$, showing that for an instable
configuration CRDSA is remaining stable much longer than
SA.

\section{Summary and Conclusions}\label{SEC:SummaryAndConclusions}
In this paper, a theoretical model for the stability of CRDSA
as representant for SIC RA schemes was developed. With
this model it is possible to draw qualitative and quantitative
conclusions about the stability of the communication channel. The
presented framework enables the estimation of the average delays
experienced in stable channel configurations. The stable CRDSA
and SA RA channels were optimized for achieving a minimum
delay, maximizing the user population while achieving a delay target
or deriving the maximum traffic generation probability for a given
user population and delay target for which the channel is stable.
Numerical results were presented which allow a direct comparison of
the performance of CRDSA and SA. These results have shown
that CRDSA does not only provide a higher throughput and lower
PLR than SA but is also capable to achieve lower delays
and higher user population and traffic generation probabilities than
SA while being stable. Finally the stability framework was
extended towards instable channel configurations of CRDSA and
makes possible to predict the average time before reaching
instability. The derived model for CRDSA was validated against
simulations and the stability behaviour of CRDSA was compared
to the one of SA for instable channels. Also here CRDSA
showed a much better performance by reaching way higher average
times before failure than SA. Besides the analysis of the
stability behaviour of a channel, the presented framework enables
the computation of the optimum design parameters, in particular
$p_r$ for which the channel either remains guaranteed stable while
minimizing the average delay or the $p_r$ for an instable channel,
which results in the desired FET time.

\ifCLASSOPTIONcaptionsoff
  \newpage
\fi



\begin{thebibliography}{10}
\providecommand{\url}[1]{#1}
\csname url@samestyle\endcsname
\providecommand{\newblock}{\relax}
\providecommand{\bibinfo}[2]{#2}
\providecommand{\BIBentrySTDinterwordspacing}{\spaceskip=0pt\relax}
\providecommand{\BIBentryALTinterwordstretchfactor}{4}
\providecommand{\BIBentryALTinterwordspacing}{\spaceskip=\fontdimen2\font plus
\BIBentryALTinterwordstretchfactor\fontdimen3\font minus
  \fontdimen4\font\relax}
\providecommand{\BIBforeignlanguage}[2]{{%
\expandafter\ifx\csname l@#1\endcsname\relax
\typeout{** WARNING: IEEEtran.bst: No hyphenation pattern has been}%
\typeout{** loaded for the language `#1'. Using the pattern for}%
\typeout{** the default language instead.}%
\else
\language=\csname l@#1\endcsname
\fi
#2}}
\providecommand{\BIBdecl}{\relax}
\BIBdecl

\bibitem{Casini2007}
E.~Casini, R.~De~Gaudenzi, and O.~Herrero, ``Contention resolution diversity
  slotted {ALOHA} ({CRDSA}): An enhanced random access scheme for satellite
  access packet networks,'' \emph{Wireless Communications, IEEE Transactions
  on}, vol.~6, no.~4, pp. 1408 --1419, April 2007.

\bibitem{Liva2010}
G.~Liva, ``A slotted aloha scheme based on bipartite graph optimization,'' in
  \emph{Source and Channel Coding (SCC), 2010 International ITG Conference on},
  Siegen, Germany, April 2010, pp. 1 --6.

\bibitem{Paolini2011}
E.~Paolini, G.~Liva, and M.~Chiani, ``High throughput random access via codes
  on graphs: Coded slotted aloha,'' in \emph{Communications (ICC), 2011 IEEE
  International Conference on}, Kyoto, Japan, June 2011, pp. 1 --6.

\bibitem{Kissling2011}
C.~Kissling, ``On the stability of contention resolution diversity slotted
  aloha ({CRDSA}),'' in \emph{(GC), 2011 Global Communications Conference},
  Houston, TX, USA, December 2011, pp. 1--6.

\bibitem{D.Bertsekas1992}
D.~Bertsekas and R.~Gallager, \emph{Data Networks}, 2nd~ed.\hskip 1em plus
  0.5em minus 0.4em\relax Prentice Hall.

\bibitem{Kleinrock1973}
L.~Kleinrock and S.~S. Lam, ``Packet-switching in a slotted satellite
  channel,'' in \emph{Proceedings of the June 4-8, 1973, national computer
  conference and exposition}, ser. AFIPS '73.\hskip 1em plus 0.5em minus
  0.4em\relax New York, NY, USA: ACM, 1973, pp. 703--710.

\bibitem{Metcalfe1976}
R.~M. Metcalfe and D.~R. Boggs, ``Ethernet: distributed packet switching for
  local computer networks,'' \emph{Commun. ACM}, vol.~19, pp. 395--404, July
  1976.

\bibitem{Capetanakis1979}
J.~Capetanakis, ``Tree algorithms for packet broadcast channels,''
  \emph{Information Theory, IEEE Transactions on}, vol.~25, no.~5, pp. 505 --
  515, September 1979.

\bibitem{Lam1974}
S.~S. Lam, ``Packet switching in a multi-access broadcast channel with
  application to satellite communication in a computer network,'' S.S. Lam,
  Department of Computer Science, University of California, Los Angeles, March
  1974, also in Tech. Rep. UCLA-ENG-7429, April 1974.

\bibitem{RioHerrero2009}
O.~del Rio~Herrero and R.~D. Gaudenzi, ``A high-performance {MAC} protocol for
  consumer broadband satellite systems,'' \emph{IET Conference Publications},
  vol. 2009, no. CP552, pp. 512--512, 2009.

\bibitem{Choudhury1983}
G.~Choudhury and S.~Rappaport, ``Diversity aloha--a random access scheme for
  satellite communications,'' \emph{Communications, IEEE Transactions on},
  vol.~31, no.~3, pp. 450 -- 457, March 1983.

\bibitem{Abramson1977}
N.~Abramson, ``The throughput of packet broadcasting channels,''
  \emph{Communications, IEEE Transactions on}, vol.~25, no.~1, pp. 117 -- 128,
  January 1977.

\bibitem{Kleinrock1975}
L.~Kleinrock and S.~Lam, ``Packet switching in a multiaccess broadcast channel:
  Performance evaluation,'' \emph{Communications, IEEE Transactions on},
  vol.~23, no.~4, pp. 410 -- 423, April 1975.

\bibitem{Liva2010a}
G.~Liva, ``Graph-based analysis and optimization of contention resolution
  diversity slotted aloha,'' \emph{Communications, IEEE Transactions on},
  vol.~59, no.~2, pp. 477--487, February 2011.

\bibitem{Gallager1963}
R.~G. Gallager, ``Low-density parity-check codes,'' MA: M.I.T. Press, 1963.

\bibitem{Abramson1970}
N.~Abramson, ``The aloha system: Another alternative for computer
  communications,'' in \emph{Proceedings of the 1970 Fall Joint Comput. Conf.,
  AFIPS Conf.}, vol.~37, Montvale, N.~J., 1970, pp. 281--285.

\bibitem{Roberts1975}
L.~G. Roberts, ``Aloha packet system with and without slots and capture,''
  \emph{SIGCOMM Comput. Commun. Rev.}, vol.~5, pp. 28--42, April 1975.

\bibitem{Gallager1985}
R.~Gallager, ``A perspective on multiaccess channels,'' \emph{Information
  Theory, IEEE Transactions on}, vol.~31, no.~2, pp. 124 -- 142, March 1985.

\bibitem{Murali1997}
R.~Murali and B.~Hughes, ``Random access with large propagation delay,''
  \emph{Networking, IEEE/ACM Transactions on}, vol.~5, no.~6, pp. 924 --935,
  December 1997.

\bibitem{Carleial1975}
A.~Carleial and M.~Hellman, ``Bistable behavior of aloha-type systems,''
  \emph{Communications, IEEE Transactions on}, vol.~23, no.~4, pp. 401 -- 410,
  April 1975.

\end{thebibliography}

\end{document}